\documentclass[aps,prl,nofootinbib,twocolumn,superscriptaddress,floatfix,10pt,a4paper]{revtex4-1}

\usepackage{graphicx}
\usepackage{amsmath}
\usepackage{xcolor}
\definecolor{blue}{rgb}{0,0,0.5}
\usepackage[colorlinks,linkcolor=blue,urlcolor=blue,citecolor=blue]{hyperref}

\begin{document}

\title{\boldmath A New Light Particle in $B$ Decays?}

\author{Filippo Sala}
\affiliation{LPTHE, UMR 7589 CNRS, 4 Place Jussieu, F-75252, Paris, France}

\author{David M. Straub}
\affiliation{Excellence Cluster Universe, Boltzmannstr.~2, 85748~Garching, Germany}

\date{\today}

\begin{abstract}\noindent
We investigate the possibility whether the tensions with SM expectations observed
in several $b \to s \ell \ell$ transitions, including hints for lepton flavour non-universality
, could be due to the decay of $B$ into a new \textit{light} resonance.
We find that qualitative agreement with the data can be obtained with
a light vector resonance dominantly decaying invisibly. 
This scenario predicts a shift in the muon anomalous magnetic moment that could
explain the long-standing discrepancy. The most stringent constraint comes
from searches for $B$ decays with missing energy
.
A striking prediction is a strong $q^2$ dependence of the lepton flavour
universality ratios $R_K$ and $R_{K^*}$ that should allow to clearly confirm
or rule out this possibility experimentally.
We also comment on the possible connection of the invisible decay product
with Dark Matter.
\end{abstract}

\maketitle

\section{Introduction}

Rare semi-leptonic $B$ decays based on the $b\to s\ell\ell$ transition are
sensitive probes of physics beyond the Standard Model (SM). In recent years,
several tensions between SM expectations and experiments have been observed,
including in particular an apparent enhancement of the angular observable
$S_5$ (or $P_5'$)\footnote{%
See \cite{Altmannshofer:2008dz,DescotesGenon:2012zf} for definitions of these observables.}
in $B^0\to K^{*0}\mu^+\mu^-$~\cite{Aaij:2013qta,Aaij:2015oid,ATLAS-CONF-2017-023},
a suppression of the branching ratios
of $B_s\to \phi\mu^+\mu^-$~\cite{Aaij:2015esa}, $B^+ \to K^{(*)+}\mu^+\mu^-$~\cite{Aaij:2014pli},
$B^0 \to K^{(*)0}\mu^+\mu^-$~\cite{Aaij:2014pli,Aaij:2016flj}, 
and a suppression of the ratio $R_K$ of dimuon to dielectron rates
in $B^+\to K^+\ell^+\ell^-$ \cite{LHCbtalk}.
Very recently, the LHCb collaboration has also revealed an apparent suppression of
the analogous ratio $R_{K^*}$ in $B^0\to K^{*0}\ell^+\ell^-$
decays \cite{Aaij:2014ora}.

The tensions in branching ratios and angular observables
could be due to underestimated theoretical uncertainties, e.g. in form factors
or from unexpectedly large non-factorisable hadronic effects.
Barring statistical fluctuations or experimental systematics, deviations from 
lepton flavour universality are instead clear indications of physics beyond the SM.
The most straightforward new physics (NP) explanation of these anomalies makes
use of a suppression of the semi-leptonic vector operator
$O_9\propto(\bar s_L\gamma^\mu b_L)(\mu\gamma_\mu\mu)$ caused by destructive
interference of the SM loop contribution and a tree- or loop-level exchange of
new heavy particles (see e.g.~\cite{Altmannshofer:2017fio} for a recent global fit to $b\to s\mu\mu$ data).
A second possibility is to have NP generate a local operator
with the quark content $\bar sb\bar cc$ \cite{Jager:2017gal}; such an effect would however be
lepton flavour universal and thus it would not explain $R_K$ nor $R_{K^*}$.
In this letter, we consider a third possiblity: a NP
effect due to new \textit{light} particles produced on shell in the $B$ decay.

Our letter is organized as follows. We first define our phenomenological setup and discuss the constraints on its parameters.
We then show how a light new particle
could solve the aforementioned tensions in $b \to s \ell \ell$ transitions, first all of them, and
then only some selected ones.
Finally, we briefly explore a possible connection with Dark Matter, and conclude.
\section{Setup}

Since several tensions have been observed in decays based on the $b\to s\mu^+\mu^-$
transition, the deviations from lepton flavour universality (LFU), $R_K$ and $R_{K^*}$,
are explained much better by models that suppress the $b\to s\mu^+\mu^-$ transition
rather than enhance the $b\to se^+e^-$ transition.
We thus require the SM amplitude to interfere destructively with the on-shell production of the new state,
in the region of dimuon invariant mass $q^2$ relevant for the $R_K$ and $R_{K^*}$ measurements.
We are then led to consider a new vector boson $V$ with mass below the $B$ meson mass,
since a new scalar would induce a negligible interference.
We anticipate that a sizeable width will be needed, since no clear peaks (besides the known QCD resonances)
have been observed in $b \to s\mu\mu$ decays.

We consider a simplified model valid at energy scales of the order of a few GeV
which, in addition to the SM states, contains a single uncoloured vector particle
$V$ with mass $m_V$ and a SM singlet lighter than $m_V/2$, either fermion or scalar.
The latter escapes the detector unmeasured and is required to allow $V$ to have
a sizable decay width without violating existing bounds. The Lagrangian of this
setup reads
\begin{equation}
\begin{split}
\mathcal{L} =  \big[(g_{bs}\,\bar{s}_L \gamma_\nu b_L + {\rm h.c.})\,
+ g_{\mu V}\,\bar{\mu} \gamma_\nu \mu
+ g_{\mu A}\,\bar{\mu} \gamma_\nu\gamma_5 \mu
\\
+ g_\chi\,\bar{\chi}\gamma_\nu \chi\big] V^\nu
+ \frac{m_V^2}{2}V^\nu V_\nu,
\end{split}
\label{eq:L}
\end{equation}
where we only allow the couplings we require, and we assume them to be real for simplicity.
For definiteness we have
chosen the light SM singlet to be a Dirac fermion $\chi$.
We further require $g_\chi\gg g_{\mu V,A}$ such that $\text{BR}(V\to\chi\chi)\simeq1$.
In this case, the decay width of $V$ is given by
\begin{equation}
\Gamma_V \simeq \frac{m_V}{12 \pi}\, g_\chi^2 \,.
\end{equation}
A large relative width requires a strong coupling $g_\chi\gtrsim 1$;
alternatively, the single field $\chi$ could be replaced by a sizeable number
of more weakly coupled $\chi_i$.

The flavour-changing $V$ coupling could arise for instance from a tree-level $\bar ttV$ coupling
permitting a penguin diagram similar to the SM $W$-top loop;
this case would also predict $b\to d$ and $s\to d$ transitions, which would
however not change our discussion.
A kinetic mixing between the photon and $V$ would generically be generated at loop level,
but with a size that would again not change the phenomenology we discuss.
The couplings of $V$ to SM particles could arise from a portal to a dark sector,
possibly justifying the large invisible $V$ width,
and/or from a low energy remnant of larger flavour symmetries at high energy.
The couplings in eq.~(\ref{eq:L}) could also be understood within an ``effective $Z'$'' framework~\cite{Fox:2011qd},
where $V$ is the gauge boson of a $U(1)'$ under which the SM is neutral: new vector-like
fermions, charged under $U(1)'$, mix with the SM ones and induce their couplings to $V$.
However, for the remainder of this letter, we will not specify a UV completion
but will instead simply work with \eqref{eq:L}. We will comment
on the case of a light scalar (as opposed to the fermion $\chi$) later on.

\section{Constraints}

An important constraint comes from the process $B\to K\chi\chi$, which has the
same experimental signature as the SM process $B\to K\nu\bar\nu$.
We perform a combination of all such independent measurements by Belle and Babar
\cite{Grygier:2017tzo,Lutz:2013ftz,Lees:2013kla,delAmoSanchez:2010bk},
and find $\text{BR}(B\to K\chi\chi) < 1.5\times 10^{-5}$ at 95\%CL.
Since the SM prediction for $\text{BR}(B\to K\nu\bar\nu)$ is roughly $0.5\times 10^{-5}$~\cite{Buras:2014fpa}, we impose the
upper bound $\text{BR}(B \to K \chi\chi) < 10^{-5}$.
Using the narrow width approximation (NWA) for $V$ and using $\text{BR}(V\to\chi\chi)\simeq1$,
the branching ratio reads
\begin{equation}
\text{BR}(B\to K\chi\chi) \simeq \tau_B~\frac{m_B^3}{64\pi m_V^2}~ \lambda^{3/2} ~[f_+(m_V^2)]^2 ~ g_{bs}^2\,,
\end{equation}
where $f_+(q^2)$ is a $B\to K$ form factor that we take from~\cite{Bailey:2015dka}
and $\lambda=1+\hat m_V^4+\hat m_K^4-2(\hat m_K^2 + \hat m_V^2 + \hat m_K^2 \hat m_V^2)$
with $\hat m_i=m_i/m_B$.
This leads to a rough upper bound
\begin{equation}
g_{bs}\lesssim  [m_V/\text{GeV}]\times0.7\times10^{-8}\,.
\label{eq:BKVbound}
\end{equation}
If $V$ is broad enough for the NWA to break down, this bound becomes weaker.

Tree-level $V$ exchange also contributes to the dispersive and absorptive parts
of the $B_s$-$\bar B_s$ mixing amplitude. However, in view of the bound \eqref{eq:BKVbound},
the resulting contributions to the mass and width differences $\Delta M_s$ and
$\Delta \Gamma_s$ are totally negligible.

A loop diagram with $V$ also gives a contribution to the anomalous magnetic moment
of the muon,
\begin{equation}
\delta a_\mu =\frac{g_{\mu V}^2\,f_V\!\left(\frac{m_\mu^2}{m_V^2}\right) -5g_{\mu A}^2\,f_A\!\left(\frac{m_\mu^2}{m_V^2}\right)}{12\pi^2}\frac{m_\mu^2}{m_V^2} ,\label{eq:g-2}
\end{equation}
where $f_{V,A}(0)=1$ and the full loop function can be found e.g.\
in~\cite{Kannike:2011ng}\footnote{%
In the notation of that paper, $f_V(x)=\frac{3}{8}(2\,F_Z(x)+G_Z(x))$, $f_A(x)=\frac{3}{40}(G_Z(x)-2\,F_Z(x))$.}.
Interestingly, for $|g_{\mu A}|<\sqrt{5}|g_{\mu V}|$, the contribution has the right sign to resolve the long-standing $(g-2)_\mu$
anomaly~\cite{Davier:2010nc},
\begin{equation}
a_\mu^\text{exp} - a_\mu^\text{SM} = (287\pm 80) \times 10^{-11}.
\end{equation}
For $m_V\gg m_\mu$ and $g_{\mu A}=0$,
bringing the measurement into agreement with the prediction at $2\sigma$ would
require
$0.004 \lesssim g_{\mu}/[m_V/\text{GeV}]\lesssim 0.007$,
while only requiring the contribution not to overshoot the measurement by more than
$3\sigma$ implies $g_{\mu}/[m_V/\text{GeV}]\lesssim 0.008$. 
As seen from \eqref{eq:g-2}, in principle the bound can be weakened or even removed by tuning $g_{\mu A}$ vs. $g_{\mu V}$.

If $g_{\mu A}\neq 0$, also a tree-level contribution to the
$B_s\to\mu^+\mu^-$ decay is generated. It can be expressed in terms of shifts in
the Wilson coefficients $C_{10}$ and $C_{P}$ (which we define as in~\cite{Altmannshofer:2012az})
\begin{align}
C_{10}^V &= \frac{g_{bs} g_{\mu A}/N}{m_{B_s}^2-m_V^2+im_V\Gamma_V},
&
C_{P}^V &= -\frac{2m_\mu}{m_V^2}C_{10}^V \,,
\end{align}
where $N=G_FV_{tb}V_{ts}^* \alpha/\sqrt{2}\pi \simeq -7.7\times10^{-10}$~GeV$^{-2}$
in the usual CKM convention
and we have neglected terms of $O(m_s/m_b)$.
For an illustrative $V$ mass of 3~GeV, the branching ratio of $B_s\to\mu^+\mu^-$
is modified by less than 20\% with respect to the SM if
$|g_{bs} g_{\mu A}|< 3\times 10^{-9}$, i.e.\ $g_{\mu A}$ could be as large as
$0.1$ if $g_{bs}$ saturates \eqref{eq:BKVbound}.

If $V$ would couple also to muon neutrinos, a strong bound would be
obtained from neutrino trident production \cite{Altmannshofer:2014pba}\footnote{%
An EW-symmetric coupling to muon neutrinos would yield, for $g_{\mu A} = 0$, to $g_{\mu V} \lesssim 5\times 10^{-3}$
at $m_V = 2.5$~GeV~\cite{Altmannshofer:2014pba}. 
}.
This constitutes a model-building challenge to constructions that yield the couplings in~(\ref{eq:L}),
but since we work in a simplified model well below the electroweak scale we take the freedom to neglect this possibility.

At one-loop level also the coupling of the $Z$ to muons, constrained
by $Z$ pole mesurements at LEP and SLD \cite{ALEPH:2005ab}, is modified by a $V$ loop. One finds
\cite{Altmannshofer:2014cfa}
\begin{equation}
\frac{g_{Z\mu\mu}}{g_{Z\mu\mu}^\text{SM}} = 1 + \frac{g_{\mu V}^2+g_{\mu A}^2}{16\pi^2} K_F\!\left(\frac{m_Z^2}{m_V^2}\right),
\end{equation}
where $K_F(x)\approx -7/2+\pi^2/3+3 \ln x-\ln^2x$ for large $x$ \cite{Haisch:2011up}.
With our choice of vector-like $V\mu\mu$ coupling, this bound
turns out to be weak due to the accidentally suppressed vector coupling of the $Z$
to charged leptons in the SM, $\frac{1}{2}-2s_w^2\approx0.04$, such that the bound
only reaches $g_{\mu V,A}\lesssim 0.5$ even for $V$ masses as low as $0.5$~GeV,
and is thus irrelevant for the further discussion.

Finally, the process $Z \to \mu^+\mu^-V$ does not lead to constraints from $Z \to 4 \mu$
searches, since those have so far been performed only for $m_{\mu\mu} > 4$~GeV~\cite{CMS:2012bw,Aad:2014wra}.
It motivates searches for $Z \to \mu^+\mu^-$+ missing energy which, to our knowledge~\cite{Olive:2016xmw},
have not been performed so far.

\section{\boldmath Resonances in \texorpdfstring{$B$}{B} decays}

Having obtained the most important constraints on the couplings of the light
vector $V$ to quarks and leptons, we can now proceed to discuss the impact on
rare exclusive semi-leptonic $B$ decays such as $B\to K\mu^+\mu^-$ and
$B\to K^*\mu^+\mu^-$. Tree-level exchange of the
resonance $V$ leads to a contribution to the amplitudes of these processes
that can be most simply expressed as a $q^2$ dependent shift of the Wilson
coefficients $C_9$ and $C_{10}$,
\begin{equation}
C_{9,10}^V = \frac{g_{bs} g_{\mu V,A}/N}{q^2-m_V^2+im_V\Gamma_V}.
\label{eq:C9}
\end{equation}

\begin{figure}[tbp]
\includegraphics[width=\columnwidth]{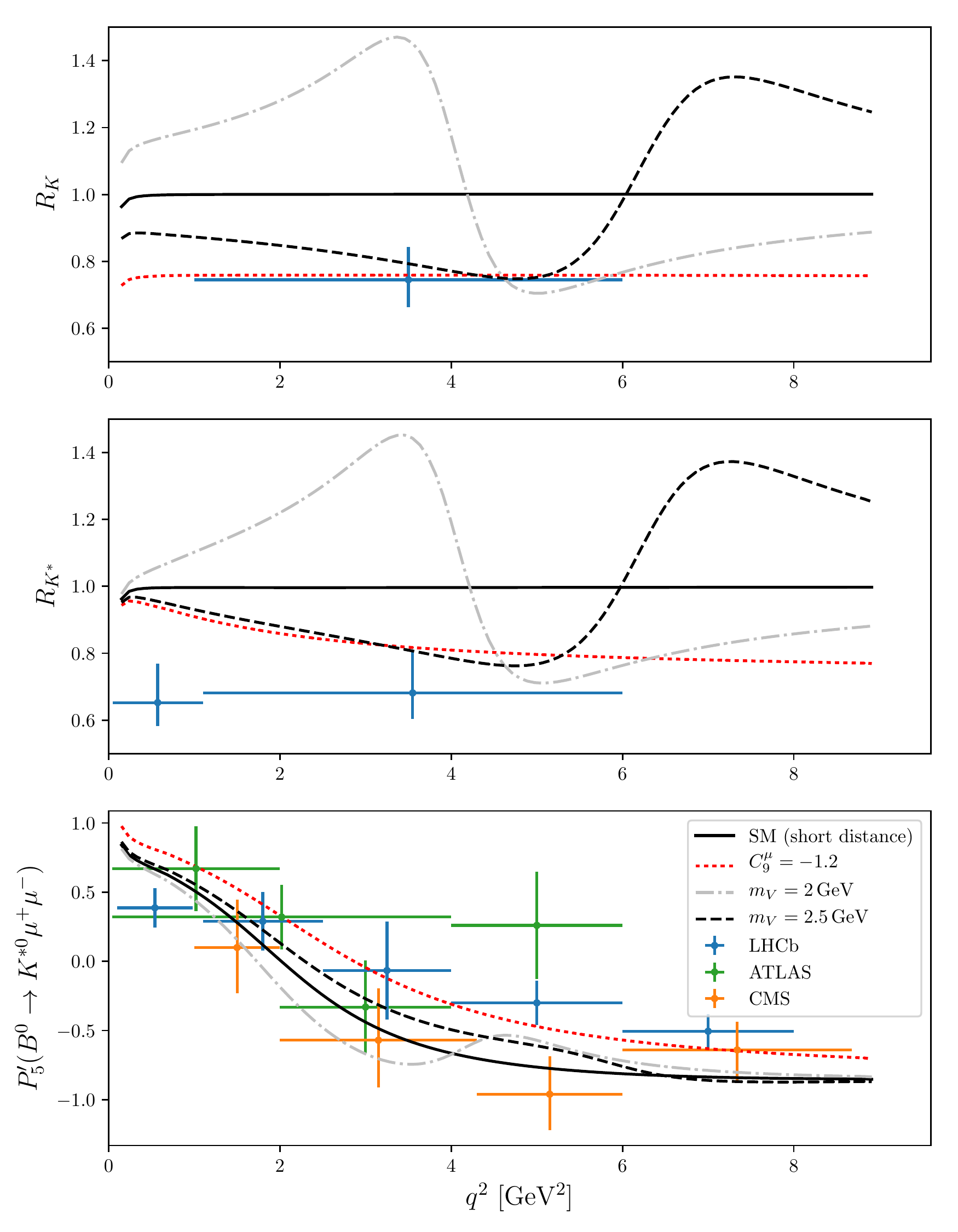}
\caption{$R_K$, $R_{K^*}$, and $P_5'$ as functions of $q^2$ for the SM,
for a heavy NP contribution to $C_9^\mu$, and for
two different scenarios with a light vector resonance, compared to experimental
results. In both light NP scenarios, the relative width of $V$ is 20\%, the bound
\eqref{eq:BKVbound} is saturated, $g_{\mu V} = 0.1$ and
$g_{\mu A}\simeq -0.44\,g_{\mu V}$  to reproduce
the experimentally observed central value of $(g-2)_\mu$.
A marked difference between the $C_9^\mu$ and the light vector contributions
persists also at values of $q^2$ higher than those shown here.}
\label{fig:q2}
\end{figure}

We are mainly interested in addressing three anomalies: the suppression of $R_K$
as well as $R_K^*$ for $q^2$ between 1 and 6~GeV$^2$
and the enhancement of $P_5'$ for $q^2$ between 4 and 6~GeV$^2$ as seen
by ATLAS and LHCb.
If the resonance is fully contained in the $q^2$ bin to be integrated over,
no significant suppression of the rate can be expected. Thus to explain $R_K$ and $R_{K^*}$
in the region from 1 to 6~GeV$^2$, a resonance with mass below 1~GeV or above
$2.5$~GeV is preferable.
If the resonance is sufficiently close in mass to the $J/\psi$ at $3.1$~GeV,
it would be very hard to detect directly. Indeed, due to the $B\to K\chi\chi$ constraint,
the branching ratio $B\to KV$ is guaranteed to always be well below the
$B\to J/\psi K$ one.
If the mass instead lies between $2.5$ and 3~\text{GeV}, a broad resonance
could not be excluded at present in view of the sizeable hadronic uncertainties
of the SM prediction in this region (cf.~\cite{Khodjamirian:2010vf,Lyon:2014hpa})
and the unknown phase of the interference
between the $J/\psi$ and the SM short-distance contribution, leaving even the
sign of the SM effect undetermined. We refrain here from a precise quantification
of the constraints on the resonance width, and we simply work with
$\Gamma/m_V \gtrsim 10\%$. This choice allows, for \textit{any} $q^2 > 6$ GeV$^2$
and for the values of the parameters preferred by the flavour anomalies,
to contain the deviation from the SM to less than 30\% or so (see e.g. Fig.~\ref{fig:q2},
analogous results hold for the other $b \to s \mu \mu$ observables).

\begin{figure*}[tbp]
\includegraphics[width=\textwidth]{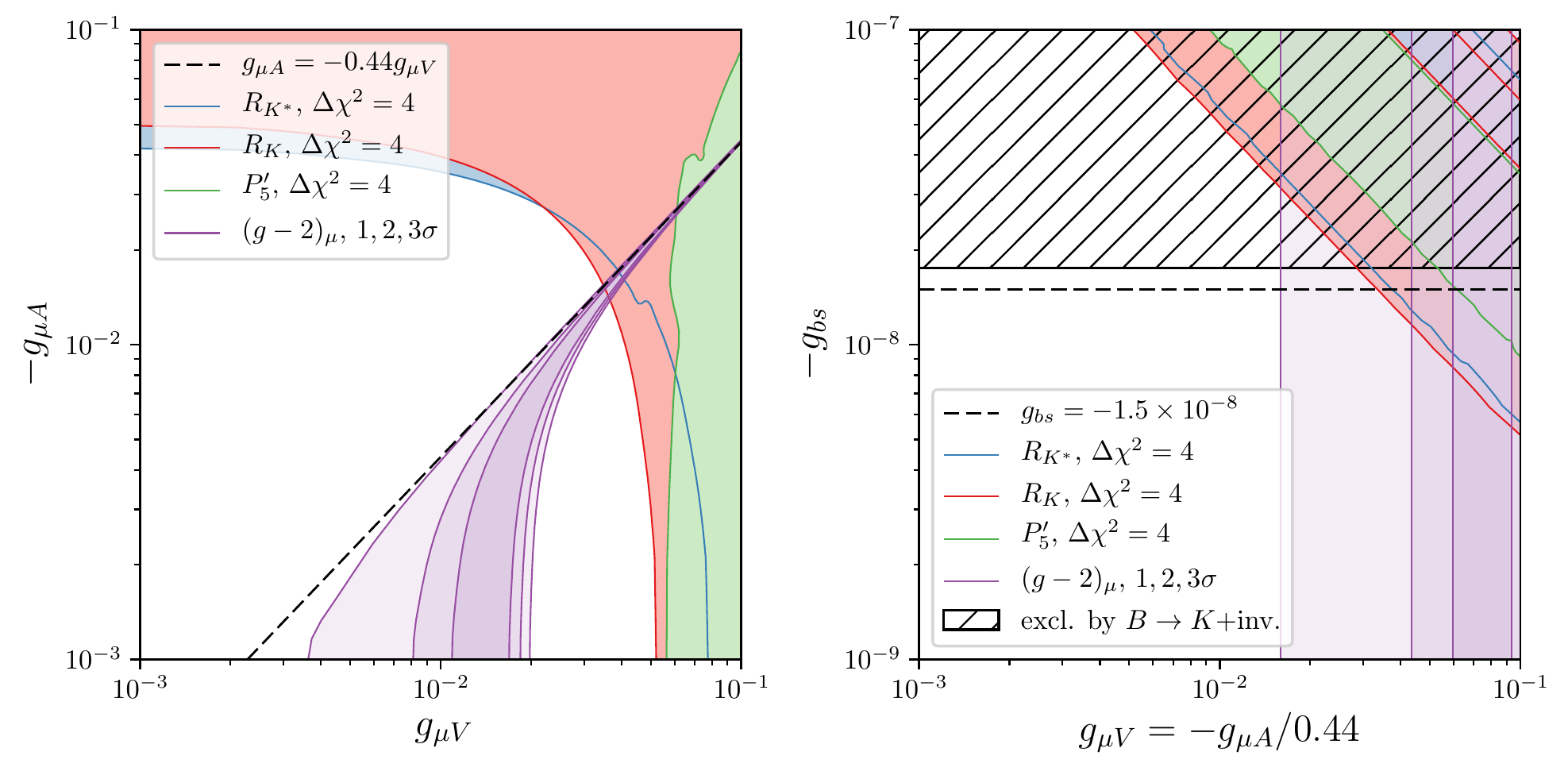}
\caption{Preferred regions from flavour anomalies and constraints for the
benchmark case of $m_V=2.5$~GeV.
Left: $g_{\mu V}$ vs. $g_{\mu A}$ fixing $g_{bs}=-1.5\times10^{-8}$.
Right: $g_{\mu V}$ vs. $g_{bs}$ fixing $g_{\mu A}=-0.44 g_{\mu V}$.}
\label{fig:gV_gA_gbs}
\end{figure*}

Having identified the potentially interesting mass and width region,
we next proceed to evaluate the effect on the observables of interest while taking into
account the constraints discussed above.
Saturating the bound \eqref{eq:BKVbound} on the flavour-changing coupling and
taking $g_{\mu A}=0$, it turns out that the effects in the $b \to s \mu \mu$ transitions
are rather small. Thus we require a mild tuning between $g_{\mu A}$ and $g_{\mu V}$
to obtain a suppression of $(g-2)_\mu$ by a factor of 3--5. 
This tuning
is apparent in Fig.~\ref{fig:gV_gA_gbs} left, where we also show the regions
preferred at $2\sigma$\footnote{To determine these regions, we have performed a fit to the
experimental data on $R_K$ and $R_{K^*}$ by LHCb and to $P_5'$ in the bins between
1 and 6~GeV$^2$ by ATLAS, CMS, and LHCb, using the \texttt{flavio} code \cite{flavio}
and the SM $P_5'$ prediction from~\cite{Straub:2015ica}.
By ``$2\sigma$'' we mean that the $\chi^2$ is improved with respect to the SM by 4.}
by $R_K$, $R_{K^*}$ and $P_5'$, and where we have saturated
the upper bound on $g_{bs}$ in eq.~(\ref{eq:BKVbound}) for definiteness.
We find that an explanation of the flavour anomalies and $(g-2)_\mu$ is possible\footnote{
The LHCb and $(g-2)_\mu$ anomalies can be both addressed also in models
with a heavy $Z'$ if new leptons are added,
see e.g.~\cite{Belanger:2015nma,Allanach:2015gkd,Altmannshofer:2016oaq}.
},
compatibly with all the other constraints, for axial and vector coupling of $V$ to muons both
different from zero and of the order of a few times $10^{-2}$.
In the ``funnel'' region\footnote{While $(g-2)_\mu$ is insensitive
to the signs of $g_{\mu A}$ and $g_{\mu V}$, we do not show their other possible relative
sign since it leads to a slightly worse fit of the $b \to s \ell \ell$ anomalies.
} $g_{\mu A} \simeq -0.44 g_{\mu V}$ the interval
\begin{equation}
10^{-9} \lesssim |g_{bs}\,g_{\mu V}| \lesssim 3\times 10^{-9}
\end{equation}
can improve all three sets of flavour observables --
$R_K$, $R_{K^*}$ and $P_5'$ -- by more than $2\sigma$ each,
as shown in Fig.~\ref{fig:gV_gA_gbs} right,
together with the region preferred $(g-2)_{\mu}$, and with the limit
$B \to K+$invisible.

To explain the enhancement of $P_5'$, but not $R_K$ and $R_{K^*}$,
a possibility is a resonance with mass just below the bin of interest,
i.e. below 2~GeV.
While the measurement of 
$B\to K^*\mu^+\mu^-$ angular observables
have been performed in relatively wide bins, a measurement 
of the differential decay rate of $B^+\to K^+\mu^+\mu^-$ in fine $q^2$ bins
has been performed in \cite{Aaij:2016cbx}. A detailed analysis of the
compatibility of our scenarios with this measurement is beyond the scope of
our analysis, here we just note that
the inclusion of a resonance with mass around 
$1.8$~GeV was found in \cite{Aaij:2016cbx} to ``marginally improve the fit quality'', while
its mean and width do not correspond to a known state.
The sign of the interference of the apparent resonance corresponds to the one
of our $m_V=2$~GeV scenario in Fig.~\ref{fig:q2}.

Finally, one could wonder whether an analogous explanation could be the reason
for the downward fluctuation in the lower $q^2$ bin of $R_{K^*}$, for example
with a resonance with couplings like in~(\ref{eq:L}) and mass $\gtrsim \sqrt{1.1}$ GeV.
Such explanation would in principle be possible,
at the price of an upward fluctuation in the second $q^2$ bin of $R_{K^*}$
(where instead a downward fluctuation is observed), and would be generically disfavoured
by the various constraints, that become stronger for smaller $V$ masses.
Therefore we do not explore this possibility further here\footnote{A very light,
narrow vector boson has been considered in
\cite{Fuyuto:2015gmk} and found not be able to reconcile the $R_K$ and $(g-2)_
\mu$ anomalies. A narrow boson even lighter than the muon was considered in
\cite{Datta:2017pfz} and found to be only viable by postulating a momentum-
dependent coupling to quarks.}.

\section{Comments on Dark Matter}
\label{sec:DM}
It is natural to speculate whether the light new particle, responsible for the invisible $V$ decay,
could be a good Dark Matter (DM) candidate for the values of the parameters that explain the flavour anomalies.
If this new particle is stable on cosmological scales, for fundamental or accidental reasons,
a straightforward possibility to avoid overclosure of the universe is that its mass be larger
than $m_\mu$.  The related annihilation cross section, in the limited mass range $m_\mu < m_{\rm DM} < m_V/2$,
is then univocally determined by the requirement to alleviate the $b\to s \mu \mu$ tensions,
and its value can for example be compared with the one needed to obtain
the DM abundance via thermal freeze-out~\cite{Steigman:2012nb}.

The average annihilation cross section of $\chi\bar{\chi} \to \mu^+\mu^-$ times relative velocity reads
\begin{equation}
\langle \sigma v \rangle
\simeq \frac{g_\chi^2}{\pi}\,\frac{g_{\mu V}^2 \big(m_\chi^2 + \frac{m_\mu^2}{2}\big)
+ g_{\mu A}^2 \big(m_\chi^2 - m_\mu^2\big)}{(m_V^2 - 4 m_\chi^2)^2}\sqrt{1-\frac{m_\mu^2}{m_\chi^2}}
\end{equation}
and yields to values in the ballpark of $10^{-(22-21)} $cm$^3$/sec
for the preferred values of $g_\chi$ and $g_{\mu V,A}$.
The annihilation of $\chi$ with $\bar{\chi}$ is then efficient in depleting the symmetric DM population,
so that either $\chi$ or $\bar{\chi}$ could be a viable DM candidate in the presence of an initial asymmetry.
Here we limit ourselves to notice this interesting property,
whose further exploration goes beyond the scopes of this letter.
The invisible stable state could also be a new light complex scalar particle $\phi$
(for a first exploration of sub-GeV scalar DM see~\cite{Boehm:2003hm}).
Its interaction with $V$, $\mathcal{L} \supset i\, g_\phi V^\nu \phi^*\partial_\nu \phi + {\rm h.c.}$,
could for example arise if the $\phi$ potential spontaneously breaks the gauge group of $V$,
thus also providing an origin for $m_V$.
The $\phi\bar{\phi}$ annihilation cross section into muons times relative velocity then reads
\begin{equation}
\langle \sigma v \rangle \simeq
v^2 \frac{g_\phi^2}{3\pi} \frac{g_{\mu V}^2 \big(m_\phi^2 + \frac{m_\mu^2}{2}\big)
+g_{\mu A}^2 \big(m_\phi^2 - m_\mu^2\big)}{(m_V^2 - 4 m_\phi^2)^2} 
\sqrt{1-\frac{m_\mu^2}{m_\phi^2}}.
\end{equation}
Like in the fermion case, and despite the $p$-wave suppression,
this cross section is large ($\langle \sigma v \rangle \approx 10^{-(24-23)}
$cm$^3$/sec at freeze-out)
for the values of the couplings preferred by the flavour anomalies
and efficiently depletes the symmetric DM population.

\section{Summary \& outlook}

In view of various deviations from SM predictions in rare $B$ decays, we
have speculated on the logical possibility of new \textit{light} particles
being responsible for these effects. Considering the case of a new light spin-1
boson, we have demonstrated that the effects can be qualitatively reproduced if
\begin{itemize}
 \item[$\diamond$] the resonance has a mass between about 2 and 3~GeV,
 in particular a mass $\gtrsim 2.5$~GeV allows for a common explanation of $R_K$,
 $R_{K^*}$ and $P_5'$;
 \item[$\diamond$] the resonance dominantly decays invisibly with a partial width that
 is of order 10--20\% of its mass.
\end{itemize}
The latter property has lead us to speculate about the invisible state as Dark Matter.
The values of the parameters that explain the flavour anomalies predict
that, if the new invisible state is heavier than $m_\mu$, it has a large pair annihilation cross section,
which is a necessary propriety 
of a would-be asymmetric DM candidate.

Independently of Dark Matter, 
this scenario makes a number of clear-cut predictions,
\begin{itemize}
 \item[$\diamond$] a strongly $q^2$-dependent deviation from $e$-$\mu$ universality in
 $B\to K\ell^+\ell^-$ and $B\to K^*\ell^+\ell^-$,
 \item[$\diamond$] a branching ratio $B\to K+\text{invisible}$ not far from present bounds
 on $B\to K\nu\bar\nu$,
 \item[$\diamond$] an effect in $(g-2)_\mu$ that can explain the observed discrepancy with the SM.
\end{itemize}
A moderate tuning at the level of a few per cent is required in the muon couplings
in order to satisfy the $(g-2)_\mu$ constraint.

Especially the first prediction --  a marked $q^2$ dependence in $R_K$
and $R_{K^*}$ -- is a unique signature that will allow to confirm or refute
this setup experimentally. We look forward to precise measurements of the
$q^2$ dependence of these observables.
\paragraph{Note added} After the first version of this paper appeared, ref.~\cite{Bishara:2017pje}
pointed out that precise measurements of the invariant mass of dimuons from Drell-Yann production
significantly constrain resonances of the kind studied in this letter.
The limits derived in~\cite{Bishara:2017pje}
reduce the parameter space where $P_5'$ can be addressed, but leave ample margin to explain
$R_K$ and $R_{K^*}$.

\subsection*{Acknowledgments}
We thank Fady Bishara, Martino Borsato, Pierre Fayet, Ulrich Haisch and
Pier Francesco Monni for interesting feedback. 
The work of DS was supported by the DFG cluster of
excellence ``Origin and Structure of the Universe''.
The work of FS is supported by the European Research Council
({\sc Erc}) under the EU Seventh Framework Programme (FP7/2007-2013)/{\sc Erc}
Starting Grant (agreement n.\ 278234 --- `{\sc NewDark}' project).

\bibliography{bibliography}

\end{document}